# Effects of high-pressure on the structural, vibrational, and electronic properties of monazite-type PbCrO$_4$

E. Bandiello[1], D. Errandonea[1,*], D. Martinez-Garcia[1], D. Santamaria-Perez[2], and F.J. Manjón[3]

[1]*Departamento de Física Aplicada-ICMUV, MALTA Consolider Team, Universidad de Valencia, Edificio de Investigación, C/Dr. Moliner 50, Burjassot, 46100 Valencia, Spain*
[2]*Departamento de Química Física I, Universidad Complutense de Madrid, MALTA Consolider Team, Avenida Complutense s/n, 28040 Madrid, Spain*
[3]*Instituto de Diseño para la Fabricación y Producción Automatizada, MALTA Consolider Team, Universitat Politècnica de Valencia, 46022 València, Spain*

**Abstract:** We have performed an experimental study of the crystal structure, lattice-dynamics, and optical properties of PbCrO$_4$ (the mineral crocoite) at ambient and high pressures. In particular, the crystal structure, Raman-active phonons, and electronic band-gap have been accurately determined. X-ray-diffraction, Raman, and optical-absorption experiments have allowed us also to completely characterize two pressure-induced structural phase transitions. The first transition is isostructural, maintaining the monoclinic symmetry of the crystal, and having important consequences in the physical properties; among other a band-gap collapse is induced. The second one involves an increase of the symmetry of the crystal, a volume collapse, and probably the metallization of PbCrO$_4$. The results are discussed in comparison with related compounds and the effects of pressure in the electronic structure explained. Finally, the room-temperature equation of state of the low-pressure phases is also obtained.

PACS numbers: 62.50.-p, 61.50.Ks, 61.05.cp, 61.50.Ah

---

[*] Author to whom correspondence should be addressed. Electronic mail: daniel.errandonea@uv.es.

I.  Introduction

The *AXO₄* monazite-type compounds form an extended family of oxides [1]. Due to some interesting physical and chemical properties, several applications for these materials are already reported and under development [1]; e.g. coatings and diffusion barriers; geochronology; luminophors, lasers, and light emitters; ionic conductors; and matrix for radioactive waste management. Monazite-type compounds crystallize in a monoclinic lattice with space group $P2_1/n$ (Z = 4) which was first reported in the framework of the Manhattan project [2]. This structure (see Fig. 1a) has been accurately described by Ni *et al.* [3], who precisely determined the structure of monazite-type phosphates. The structural arrangement is based on the nine-fold coordination of the *A* cation and the four-fold coordination of the *X* cation. Monazites exist in Nature and are important accessory minerals in granitoids and rhyolites, and because of their incorporation of rare-earth elements they can effectively control the rare-earths distribution in igneous rocks [4]. In addition, they are a common accessory mineral in plutonic and metamorphic rocks. Therefore, the knowledge of the high-pressure (HP) behavior of monazites is very relevant not only for technological applications, but also for mineral physics, chemistry, and for petrology studies [5]. Monazite-type phosphates have been consequently studied under compression [6, 7] being the crystalline structure stable up to approximately 30 GPa. Cation substitution has been shown to have a relevant influence on transition pressures in *AXO₄* oxides related to monazite. In particular, magnetic cations, like Cr, reduce considerably the transition pressures in zircon-type oxides [8, 9], favoring the study of HP phases. In this work, to gain further understanding of the structural properties of monazite-type oxides, HP x-ray diffraction, Raman, and optical-absorption experiments on monazite-type $PbCrO_4$ (the mineral crocoite) up to 18 GPa are reported. We detected the occurrence of two structural changes and characterized the structure of the HP phases. The equation of state (EOS) for different structures is presented too. Finally, lattice-dynamics properties and the electronic structure of $PbCrO_4$ are studied.

**II. Experimental details**

Experiments in $PbCrO_4$ at room temperature (RT) were performed on samples obtained from natural crocoite minerals provided by Excalibur Mineral Company and collected at the Red Lead Mine, Dundas, Tasmania, Australia (see Fig. 1b). Electron microprobe analysis was performed to determine the impurities present in the natural crystal. The only detected impurity was Fe (0.06%). Crystals were translucent with a red-orange color (see Fig. 2) and shape of long thin prisms. X-ray diffraction (XRD) on powder samples finely ground from the crystals confirmed that the crystal structure is of the monoclinic monazite type.

Ambient- and HP-powder angle-dispersive x-ray diffraction (ADXRD) measurements were carried out with an Xcalibur diffractometer (Oxford Diffraction Limited). XRD patterns were obtained on a 135 mm Atlas CCD detector placed 110 mm from the sample using $K_{\alpha 1}:K_{\alpha 2}$ molybdenum radiation. The x-ray beam was collimated to a diameter of 300 µm. HP measurements on $PbCrO_4$ powder were performed in a modified Merrill–Bassett diamond-anvil cell (DAC) up to 13 GPa. The diamond anvils used have 500 µm culets. The same set-up was used previously to successfully characterize the HP phases of related oxides in the same pressure range [10, 11]. The $PbCrO_4$ powder was placed in a hole with a diameter of 200 µm drilled in a stainless-steel gasket, previously pre-indented to a thickness of 60 µm. The observed intensities were integrated as a function of 2θ in order to give conventional one-dimensional diffraction profiles. The CrysAlis software, version 171.33.55 (Oxford Diffraction Limited), was used for the data collection and the preliminary reduction of data. The indexing and refinement of the powder patterns were performed using the POWDERCELL [12] and FULLPROF [13] program packages.

Raman experiments were performed in small crystals cleaved from the natural crystals of $PbCrO_4$. HP experiments were carried out in a membrane-type DAC equipped with 500 μm anvils. Measurements were performed in the backscattering geometry using a 632.8 nm HeNe laser and a Horiba Jobin Yvon LabRAM high-resolution ultraviolet (UV) microspectrometer in combination with a thermoelectric-cooled multichannel CCD detector with spectral resolution below 2 $cm^{-1}$ [14]. For optical-absorption studies we used 20 μm thin platelets cleaved from the natural crystals. Measurements in the UV–visible–near-infrared range were done in an optical setup, which consisted of a deuterium lamp, fused silica lenses, reflecting optics objectives, and an UV-visible spectrometer [15]. For HP studies the samples were loaded in the same membrane-type DAC. Fig. 2 shows a crocoite crystal loaded into the DAC. In Raman and absorption experiments (as well as in XRD studies), a 4:1 methanol-ethanol mixture was used as a pressure-transmitting medium [16, 17]. Ruby chips evenly distributed in the pressure chamber were used in all experiments to measure the pressure by the ruby fluorescence method [18].

### III. Results and discussion

#### A. Ambient pressure characterization

A powder XRD pattern collected at ambient conditions is shown in Fig. 3a. It confirmed our samples have a crystalline structure with space-group symmetry $P2_1/n$. After a Rietveld refinement of this diffraction pattern collected outside the DAC, the following structural parameters for $PbCrO_4$ were obtained: $a = 7.098(7)$ Å, $b = 7.410(7)$ Å, $c = 6.779(7)$ Å, and $\beta = 102.4(2)$ °. The structure has four formula units per unit cell ($Z = 4$) and the unit-cell volume is 348.2(9) $Å^3$. The refinement residuals are $R^2_F = 1.58$

%, $R_{WP}$ = 2.46 %, and $R_P$ = 1.75 % for 210 reflections. These values and the refinement shown in Fig. 3a illustrate the quality of the structural solution. The obtained values agree within 1% with most unit-cell parameters reported in the literature [1]. The atomic positions, obtained for the structure, are summarized in Table I. According with these parameters, there are three different Cr-O bond distances in the slightly distorted $CrO_4$ tetrahedra: 1.6508 Å (x2), 1.6627 Å, and 1.6674 Å. On the other hand the $PbO_9$ polyhedral units are composed by eight short Pb-O bonds and a long Pb-O bond: 2.5628 Å, 2.5761 Å, 2.5902 Å, 2.6205 Å, 2.6310 Å, 2.6440 Å, 2.6478 Å, 2.7412 Å, and 3.0620 Å. There is also an isolated tenth neighboring O atom around Pb located at 3.3826 Å, which could become relevant upon compression favoring a coordination increase. The reported distances agree well with those previously published by Effenberger and Pertlik [19].

Figure 3b shows a Raman spectrum measured for $PbCrO_4$ at ambient conditions. Due to the distinct structural units within monazite ($PbO_9$ and $CrO_4$), to understand the lattice vibrations of it, we may regard the structure to a good approximation as being composed of two sub-lattices containing separately Pb and $CrO_4$ molecules [20]. Therefore, the vibrational modes of crocoite can be classified as either internal or external modes of the $CrO_4$ unit. The external modes correspond either to a pure translation (T) or to a pure rotation (R) of the $CrO_4$ molecule; while the internal modes can be decomposed into four types of motion ($v_1$, $v_2$, $v_3$, and $v_4$), which correspond to stretching ($v_1$ and $v_3$) and bending ($v_2$ and $v_4$) vibrations. Group theoretical considerations lead to the following vibrational representation at the Γ point for monazite in the standard notation: $\Gamma = 18 A_g + 18 B_g$, whose classification into internal and external modes yields $\Gamma = A_g(6T, 3R, v_1, 2v_2, 3v_3, 3v_4) + B_g(6T, 3R, v_1, 2v_2, 3v_3, 3v_4)$. Usually in monazites, no more than twenty-two modes have been detected [20].

Previously for PbCrO$_4$ a maximum of ten modes has been reported. In the present experiment we found twenty-six Raman-active phonons which are summarized in Table II and compared with previous results. Other four modes, making a total of thirty, are detected in HP experiments due to mode splitting. It can be seen there that our results agree well with previous studies [21 – 23] being the agreement quite good with the results reported by Frost [23]. Among the previous known modes there is only one that we did not find. The mode reported at 36 cm$^{-1}$ by Wilkins [21] cannot be seen due to the cut-off of our edge filter. On the other hand, we observed twenty modes that were previously not reported, most of them in the low-frequency part of the spectrum. From Fig. 3b we can say that the Raman spectrum of crocoite has five high-frequency modes, like monazite phosphates [20]. They are most likely due to stretching motions within the CrO$_4$ molecule. The most intense mode is the one at 840.2 cm$^{-1}$. By analogy with other monazites we think it is reasonable to assign this mode to the $\nu_1$ symmetric stretching vibration. In addition, the Raman spectrum shows ten modes in the middle wavenumber region (326 – 480 cm$^{-1}$), which can be assigned to bending vibrations of the CrO$_4$ units [23]. Finally, we detected fifteen vibrations in the low wavenumber region of the spectrum, which are originated by external (lattice) vibrations.

The absorption coefficient ($\alpha$) of PbCrO$_4$ at ambient conditions is shown in Fig. 3c. Given the thickness of the studied sample and the stray-light level of our spectroscopic system, the highest measurable value of the absorption coefficient is of the order of 2000 cm$^{-1}$, which is a typical value for a low-energy tail of a direct-absorption edge [24]. The absorption spectrum shows a steep absorption, characteristic of a direct band-gap, plus a low-energy absorption band which overlaps partially with the fundamental absorption. This low-energy absorption band has been previously observed in related metal tungstates and seems to be related to the presence of defects or impurities [24]. Regarding the steep absorption edge, we found it exhibits an exponential dependence on the photon energy following Urbach's law [25]. Therefore, in order to determine $E_g$, we have analyzed the measured absorption spectrum assuming $\alpha = A_0\, e^{-\frac{E_g - h\nu}{E_u}}$. In this equation $E_u$ is Urbach's energy, which is related to the steepness of the absorption tail, and $A_0 = k\sqrt{E_u}$ for a direct band-gap [26], being $k$ a characteristic parameter of each material. Fig. 3c illustrates the quality of the fit we got for our data using this model. As can be seen, the agreement of the fit with the experiments is quite good. From it we obtained $E_g$ = 2.3 eV and $E_u$ = 60 meV. The band-gap energy is comparable with those obtained from diffuse-reflectance measurements performed in powder samples (2.2 – 2.4 eV) [27, 28].

**B. High-pressure XRD measurements**

The *in situ* ADXRD data of PbCrO$_4$ measured at different pressures are shown in Fig. 4. The x-ray patterns could be indexed with the monazite structure, stable at ambient conditions, up to 3.25 GPa. At this pressure the splitting of several peaks is observed (see arrows in the plot). This phenomenon is enhanced at higher pressures (see figure), but the patterns can be assigned to a monoclinic structure with the same symmetry than monazite. Upon further compression gradual changes occur in the diffraction patterns from 5.2 to 9.1 GPa. These changes can be attributed to the onset of a phase transition at 6.1 GPa, which is fully completed at 9.1 GPa; i.e. two phases coexist in this pressure range. Upon further compression no additional changes are found in the patterns up to the highest pressure reached in the experiments. Finally, a diffraction pattern of the monazite structure is recovered on decompression, indicating that structural changes are reversible.

From the diffraction patterns collected at different pressures we extracted the pressure evolution of the unit-cell parameters of PbCrO$_4$ up to 8.1 GPa. Results are summarized in Fig. 5. There it can be seen than compression is non isotropic, being the *a* axis the most compressible one and the c axis the least one. Similar behavior was previously found for the thermal expansion [29]. This anisotropic behavior is caused by the fact that monazite-type PbCrO$_4$ is composed by chains of alternating PbO$_9$ and CrO$_4$ polyhedra aligned along c axis, while a void space exists among polyhedral units in the other directions [3]. In Fig. 5, it can be also seen than there is a slope change in the pressure evolution of different parameters at 3.25 GPa. This change is clearly noticeable for the c axis and the β angle. It indicates the occurrence of an isostructural transition which involves a strong structural distortion of the monoclinic structure of PbCrO$_4$, but not a change in the space-group symmetry of the crystal. As we will see latter, this structural distortion, which gradually increases the Pb coordination from 9 to 10, has consequences on the optical properties of PbCrO$_4$ and can be correlated with changes observed in the Raman spectra. On the other hand, as pressure increases the monoclinic structure tends to become more symmetric, as observed in monazite-type phosphates [6]. From the unit-cell data we also obtained the pressure evolution of the unit-cell volume, which is shown in Fig. 5. Notably, we cannot detect any subtle change in the volume or the volume compressibility around 3.25 GPa, thus indicating that the isostructural phase transition is likely of second-order displacive type. We fit the volume data of Fig. 5 with a Birch-Murnaghan EOS [30]. Since we have only nine data points we fixed the volume at ambient pressure to 348.2 Å$^3$, and the pressure derivative of the bulk modulus to 4. Thus we obtained a bulk modulus (B$_0$) of 57(3) GPa. This EOS describes the P-V relation at RT for both monoclinic structures. The value of B$_0$ indicates that monazite chromates are much more compressible than monazite phosphates which have bulk moduli larger than 100 GPa [6].

We have also made an attempt to identify the structure of the high-pressure phase of PbCrO$_4$ observed beyond 9.1 GPa. The reduction of the number of Bragg reflections suggests an increase of the symmetry of the crystal. Based upon this fact and the crystal-chemistry arguments proposed in Ref. [31], we have considered several potential structures previously found in *AXO$_4$* compounds. In particular, we evaluated the low- and high-pressure structures already observed or predicted for arsenates [9], phosphates [6], vanadates [32], germanates [33], silicates [34], sulphates [35], chlorates [36], molybdates [10], and tungstates [31, 37]. Among these structures, the barite-type structure found in PbCrO$_4$ at high temperature [38] has to be excluded since clearly does not explain the reflexions we observed in diffraction patterns beyond 9.1 GPa. After a deep analysis, we found that these diffraction patterns could be indexed with an orthorhombic cell. For the pattern measured at 12 GPa we obtained the following lattice constants: $a = 6.95(6)$ Å, $b = 6.11(6)$ Å, and $c = 6.63(6)$ Å [V = 282(8) Å$^3$, Z = 4]. By extrapolating the low-pressure volume to 12 GPa, we estimated that the orthorhombic HP structure implies a volume change of about 5% at the transition. The systematic absences in the indexed lattice planes are consistent with symmetry elements of the space group *P2$_1$2$_1$2$_1$*. Therefore, the structure of the HP phase is related to the one found in BaSO$_4$ [35] upon compression and to the structure of CaSeO$_4$ [39]. The appearance of this structure at HP is fully consistent with crystal-chemistry arguments [31] and with the idea that pressure should induce a cation coordination increase. The structure proposed for the HP phase is basically a strong distortion of barite formed by CrO$_4$ tetrahedra and PbO$_{12}$ polyhedra. As we will show, its appearance is coherent with changes observed in Raman and optical experiments.

## C. High-pressure Raman measurements

Figure 6 shows RT Raman spectra of PbCrO$_4$ at different pressures up to 13 GPa. At pressures as low as 1 GPa we observed the splitting of several Raman modes of monazite. This is caused by the non-isotropic compressibility of the crystal. Also beyond 2.5 GPa some changes can be seen in the low-frequency part of the spectrum due to the hardening of modes that were originally located close to 30 cm$^{-1}$ (the cut-off of our set-up). In particular, there is a mode that starts to be seen at 2 GPa, being the extrapolated ambient pressure frequency 41.6 cm$^{-1}$. In a similar way to these modes, most of the Raman modes have a positive shift upon compression. The pressure evolution for the phonons of the low-pressure phase is summarized in Fig. 7. It is noteworthy that there is one mode near 326.9 cm$^{-1}$ which clearly softens under pressure. There are other two modes in the low-frequency region at 80.9 and 72.7 cm$^{-1}$ which also softens upon compression. The presence of these modes could be related to structural instabilities induced upon compression which trigger the phase transition detected by x-ray diffraction experiments [40]. Up to 4.5 GPa we observed a similar behavior, with no qualitative changes in the Raman spectrum. However, there are two facts that deserve to be remarked. At least six modes show a non-linear behavior (see Fig. 7), changing the frequency evolution upon compression beyond 2.5 GPa. And three weak extra modes are detected beyond 3.5 GPa around 930, 350, and 60 cm$^{-1}$. Upon further compression, at 5.3 GPa we found very clear changes that we associated to the isostructural transition detected in diffraction experiments beyond 3.25 GPa (see Fig. 6). In particular, it is noticeable the decrease of the Raman signal and the increase of the numbers of phonons in the high-frequency region and the decrease of phonons in the intermediate region (see Fig. 7). The increase of high-frequency modes can be caused by a distortion of the CrO$_4$ tetrahedra related to the strong distortion of the monoclinic structure induced after the isostructural transition. It is also interesting to note that the extra peaks detected from 3.5 to 5.3 GPa correlate well with phonons of the HP isostructural phase. So their presence and the evolution change of non-linear modes can be hints of the onset of the isostructural transition. From 5.3 to 10.1 GPa, there are no important changes in the Raman spectrum. Only the gradual appearance of weak Raman modes in the low-and intermediate-frequency region can be highlighted. The gradual appearance of peaks can be associated to the phase coexistence detected by diffraction experiments. Up to twenty-three modes are detected in the HP phase. They are summarized in Table III. Upon further compression, at 11.6 GPa the intensity of the Raman signal suddenly drops and the number of Raman modes is reduced so we have not attempted an analysis of the Raman modes of this phase. These changes are indicative of the completion of the second transition. In particular, the decrease of number of modes is consistent with the symmetry increase of the crystal that takes place at the monoclinic-orthorhombic transition. At higher pressure we did not observe any substantial change in the Raman spectrum up to 17.9 GPa (the highest pressure reached in our experiments). Upon decompression the observed changes are reversible as can be seen in Fig. 6.

From data of Fig. 7 we calculated for the low-pressure coefficient ($d\omega/dP$) and the Grüneisen parameter $\gamma = (B_0/\omega_0) \cdot (d\omega/dP)$, where $\omega_0$ is the phonon frequency at ambient pressure. To calculate $\gamma$ we assumed $B_0 = 57$ GPa, as obtained from our XRD data. The obtained parameters are summarized in Table II. Also the pressure coefficient was calculated for the Raman modes of the HP monoclinic phase (see Table III). In Table II it can be seen that all internal stretching modes (those at high frequency) have a similar pressure coefficient with the exception of the most intense mode which has a pressure coefficient about half than that of the other stretching modes. In addition, the modes with the highest pressure coefficient are the internal bending mode with the highest frequency (479.3 cm$^{-1}$) and some of the lattice modes (see Table II). The presence of soft modes (see Table II) and non-linear modes is corroborated by the fitting of the data reported in Fig. 7. For the HP phase we found pressure coefficients of the same order and also the existence of a soft mode in the low-frequency region (see Table III). The fact that modes with similar frequencies in both phases have similar pressure coefficients is related to the structural similarity between the two isostructural phases.

**D. High-pressure optical absorption measurements**

HP absorption experiments were performed in four independent samples which showed a similar behavior. In Fig. 8a, we show the absorption spectra of PbCrO$_4$ at selected pressures. There it can be seen that upon compression the absorption edge gradually red-shifts up to 2.9 GPa. At 3.5 GPa, an abrupt shift is detected, which produce the color change from red-orange to burgundy-red (see Fig. 2), indicating the occurrence of a band-gap collapse. We associated this change to the isostructural transition detected at similar pressure in diffraction and Raman experiments. Beyond 3.5 GPa, the absorption spectra also resemble those of a direct band-gap semiconductor. The pressure evolution of the absorption edge is also towards low energy. When approaching 12 GPa the formation of defects in the PbCrO$_4$ crystal is detected and at this pressure the crystal suddenly becomes black impeding the performance of optical measurements. This indicates the occurrence of a second transition, probably to a metallic phase. Note that both changes in optical properties occur at similar pressures than structural and vibrational changes, suggesting a correlation between all phenomena. Upon pressure release from pressures smaller than 12 GPa, changes are reversible (as structural ones). When releasing compression from pressures higher than 12 GPa changes in color appear to be reversible, but the presence of large number of defects does not allow the performance of accurate optical measurements (diffuse light contaminates transmission).

In order to qualitatively analyze the pressure effects on band-gap, we assumed the low- and high-pressure monoclinic phases of PbCrO$_4$ have a fundamental direct band-gap. Using the same method employed to determine $E_g$ at ambient conditions we obtained the pressure dependence of $E_g$. This method has proven to be accurate to determine the pressure effects of $E_g$ in related compounds [15, 41]. Fig. 8b shows the variation of the $E_g$ versus pressure up to 12 GPa. There is a linear decrease in the band-gap energy with increasing pressure up to 3 GPa, being the pressure coefficient -5 meV/GPa. From 3 GPa to 3.5 GPa $E_g$ abruptly changes from 2.15 eV to 1.75 eV. Beyond this pressure $E_g$ decreases upon compression in a linear way with a pressure coefficient of -4.5 meV/GPa. Variations of the Urbach energy, $E_U$ with pressure are comparable to the error of this parameter, so no conclusion on its pressure behavior can be extracted.

Based upon present knowledge of the electronic structure of monazite PbCrO$_4$ at atmospheric pressure, a qualitative approach towards the understanding of the presented results is suggested in the following. According to Stoltzfus [28], in PbCrO$_4$ the main contribution at the bottom of the conduction band results from the antibonding interaction between the Cr *3d* orbitals and the O *2p* orbitals, while the upper portion of the valence band results primarily from the interaction between Pb *6s* orbitals and O *2p* orbitals. Since the space-group symmetry permits mixing of the Pb *6s* and Cr *3d* orbitals, a minimal contribution from the Pb *6s* orbitals is observed at the bottom of the conduction band. A schematic diagram of the band structure of PbCrO$_4$ is given in Fig. 9. It has many similarities with the band structure of scheelite-type PbWO$_4$ [28]. Therefore, by analogy to PbWO$_4$, we believe than under compression, due to the increase of the crystal field, Pb *6s* states shift towards high energies faster than the Cr *3d* states [42]. This causes a reduction of the energy difference between the bottom of the conduction band and the top of the valence band inducing the $E_g$ reduction we observed up to 3 GPa.

On the other hand, the collapse of $E_g$ observed at 3.5 GPa could be caused by the isostructural change we found at similar pressures. While the structural changes do not affect the global symmetry of the crystal, the crystal structure is highly distorted, affecting probably also Cr-O and Pb-O bond distances and/or cation coordination (indeed Pb coordination increases from 9 to 10). This fact is also reflected in Raman experiments as commented. As in the case of $PbWO_4$ [15, 41], these changes of the crystalline structure should be directly reflected in the electronic structure of $PbCrO_4$, producing the collapse of $E_g$ that we observed at 3.5 GPa. These arguments provide a plausible explanation to the phenomena we observed. However, *ab initio* band-structure calculations at different pressures, like those performed in $EuWO_4$ [43], would be needed to confirm our hypothesis and the metallization of $PbCrO_4$ that apparently occurs at the monoclinic-orthorhombic transition. Metallization can be also confirmed by high-pressure transport [44] and dielectric studies [45].

### IV. Concluding Remarks

In this work we reported an experimental study of the structural, lattice-dynamics, and electronic properties of $PbCrO_4$ at ambient pressure and under compression. Experiments allowed us to accurately determine the crystal structure, Raman spectraum, and electronic band-gap of monazite-type $PbCrO_4$ (the mineral crocoite). High-pressure studies indentified the occurrence of two phase transitions. The first one occurs at 3.5 GPa, it is isostructural, and induces a band-gap collapse of 0.4 eV. The second one occurs around 9 - 12 GPa (depending upon the experimental technique). It is associated to an enhancement of the symmetry of the crystal and to a probable pressure-induced metallization. In addition, the pressure evolution of Raman modes is reported for the low- and high-pressure monoclinic phases, and the pressure dependence of unit-cell parameters and band-gap for the two isostructural monoclinic phases. Compression of these phases is highly non-isotropic. The room-temperature equation of state of the monoclinic phases is reported. Finally, the pressure evolution of the electronic band-gap is explained using known band structure models.

**Acknowledgments**

We acknowledge the financial support of the Spanish MCYT through Grants MAT2010-21270-C04-01/04 and CSD2007-00045. Financial support from the Spanish MICCIN under the Project No. CTQ2009-14596-C02-01 is also acknowledged, as well as from Comunidad de Madrid and European Social Fund: S2009/PPQ-1551 4161893 (QUIMAPRES).

**Figure captions**

**Figure 1:** (Top) Schematic view of the crystal structure of monazite-type $PbCrO_4$. (Bottom) Picture of part of the mineral from which samples were extracted.

**Figure 2:** (Top) $PbCrO_4$ crystal loaded in a DAC at ambient pressure. (Bottom) $PbCrO_4$ crystal loaded in a DAC at 3.5 GPa. The color change is indicative of the band-gap collapse.

**Figure 3:** (a) XRD pattern collected from $PbCrO_4$ at ambient conditions. Dots: measured pattern. Solid line: refined profile. Dotted line: residuals. Ticks: calculated positions for Bragg reflections. (b) Raman spectrum of $PbCrO_4$ at ambient conditions. At wavelengths smaller than 600 cm$^{-1}$ the spectrum has been magnified five times to facilitate peak identification. The ticks indicate the position of identified phonons. (c) Ambient conditions absorption spectra of $PbCrO_4$ showing the fit used to determine $E_g$. Dots: experiment. Line: fit.

**Figure 4:** Selection of room-temperature ADXRD data of $PbCrO_4$ at different pressures up to 12 GPa. In all diagrams the background was subtracted. Pressures are indicated in the plot. (r) denotes a diffraction pattern collected after pressure release. In the ADXRD patterns at 0.55 and 12 GPa we show with ticks the calculated position of Bragg peaks. Arrows indicate the appearance of peaks.

**Figure 5:** Pressure evolution unit-cell parameters and unit-cell volume. Symbols: data extracted from experiments. Lines: fits to data. For the volume the EOS is represented.

**Figure 6:** Raman spectra of $PbCrO_4$ at selected pressures.

**Figure 7:** Pressure dependence of the Raman mode frequencies in $PbCrO_4$. Symbols: data taken from experiments. Solid lines: fits to experimental data used to determine dω/dP. Solid and empty symbols were used to facilitate mode identification. Circles: low-pressure phase. Squares: high-pressure phase.

**Figure 8:** (a) Absorption spectra at representative pressures for PbCrO$_4$. (b) Pressure dependence of the band-gap energy for PbCrO$_4$. Different symbols correspond to independent experiments. Solid symbols: upstroke. Empty symbols: downstroke.

**Figure 9:** Schematic diagram of the band structure of monazite PbCrO$_4$.

**Table I:** Atomic coordinates for PbCrO$_4$ obtained from powder diffraction at ambient conditions.

| Atom | Site | x | y | z |
|------|------|-----------|-----------|------------|
| Pb   | 4e   | 0.2247(3) | 0.1515(2) | 0.4044(5)  |
| Cr   | 4e   | 0.1984(2) | 0.1643(2) | 0.8845(9)  |
| O$_1$ | 4e  | 0.2561(3) | 0.0047(1) | 0.0568(1)  |
| O$_2$ | 4e  | 0.1201(2) | 0.3415(4) | -0.0057(1) |
| O$_3$ | 4e  | 0.0274(1) | 0.1047(2) | 0.6858(7)  |
| O$_4$ | 4e  | 0.3859(4) | 0.2152(3) | 0.7872(8)  |

**Table II:** Raman frequencies at ambient conditions ($\omega_0$) compared with literature, pressure coefficients at ambient pressure, d$\omega$/dP, and Grüneisen parameters $\gamma$. The asterisks denotes modes extrapolated to ambient pressure from HP data.

| Suggested assignments | $\omega_0$ [cm$^{-1}$] | d$\omega$/dP [cm-1/GPa] | $\gamma$ | $\omega_0$ [cm$^{-1}$] | $\omega_0$ [cm$^{-1}$] | $\omega_0$ [cm$^{-1}$] |
|---|---|---|---|---|---|---|
| | This work | | | Ref. [21] | Ref. [22] | Ref. [23] |
| Lattice Modes | | | | 36 | | |
| | 41.6* | 0.7 | 0.96 | | | |
| | 45 | 1.0 | 1.27 | | | |
| | 57 | 3.7 | 3.70 | | | |
| | 61.9* | 1.6 | 1.47 | | | |
| | 72.7 | -0.2 | -0.16 | | | |
| | 80.9* | -1.8 | -1.27 | | | |
| | 83.4 | 4.2 | 2.87 | | | |
| | 94.9 | 0.4 | 0.24 | | | |
| | 99* | 2.8 | 1.61 | | | |
| | 109.6 | 4.2 | 2.18 | | | |
| | 115.5 | 6.8 | 3.35 | | | |
| | 135.5 | 5.5 | 2.31 | | | 135 |
| | 148.9 | 5.2 | 1.99 | | | |
| | 178.4 | 7.7 | 2.46 | | | |
| | 185 | 4.3 | 1.32 | | | 184 |
| Bending of CrO$_4$ | 326.9 | -0.5 | -0.09 | 326 | 325 | 327 |
| | 338 | 1.8 | 0.30 | 336 | 337 | 339 |
| | 346.7 | 1.5 | 0.25 | | 347 | |
| | 359.3 | 1.0 | 0.16 | 359 | 358 | 360 |
| | 378.1 | 2.8 | 0.42 | 377 | 377 | 379 |
| | 401.7 | 0.8 | 0.11 | 400 | 400 | 402 |
| | 407.2 | 3.7 | 0.52 | | | |
| | 439.9 | 3.9 | 0.51 | | | |
| | 450.7 | 2.2 | 0.28 | | | |
| | 479.3 | 5.4 | 0.64 | | | |
| Stretching of CrO$_4$ | 801 | 4.3 | 0.31 | | | |
| | 825 | 3.3 | 0.23 | 825 | 823 | 825 |
| | 840.2 | 1.9 | 0.13 | 838 | 839 | 841 |
| | 855.7 | 3.5 | 0.23 | 853 | 854 | 856 |
| | 879 | 3.2 | 0.21 | | | |

**Table III:** Raman frequencies (ω) at 6.4 GPa and linear pressure coefficients (dω/dP) for the HP monoclinic phase.

| ω [cm$^{-1}$] | dω/dP [cm$^{-1}$/GPa] | ω [cm$^{-1}$] | dω/dP [cm$^{-1}$/GPa] |
|---|---|---|---|
| 44 | 0.9 | 370.4 | 0.9 |
| 58.7 | 0.4 | 386 | 1.0 |
| 70.2 | 0.7 | 466 | 0.8 |
| 85 | -0.8 | 746.4 | 0.4 |
| 101 | 4 | 815.1 | 1.0 |
| 116 | 3.7 | 833.6 | 1.4 |
| 139 | 1.8 | 847 | 1.5 |
| 168 | 3.4 | 857 | 3.4 |
| 200 | 4.0 | 916 | 1.6 |
| 217 | 8.2 | 927.9 | -0.3 |
| 338 | 2.9 | 933 | 1.0 |
| 351 | 0.9 | | |

**Figure 1**

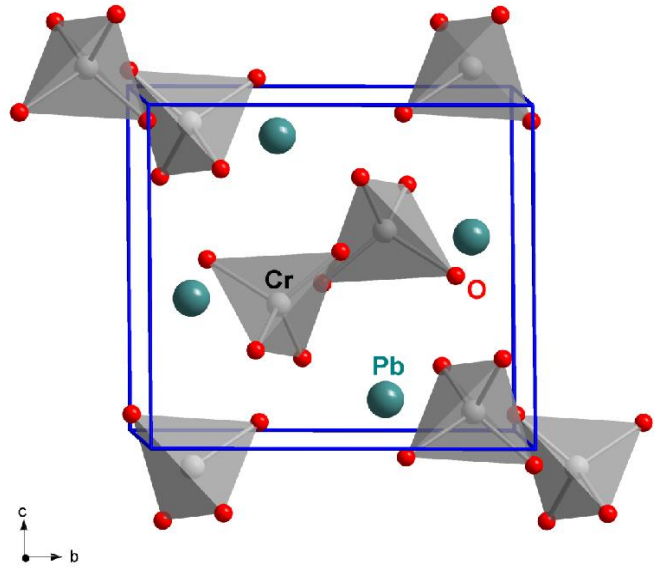

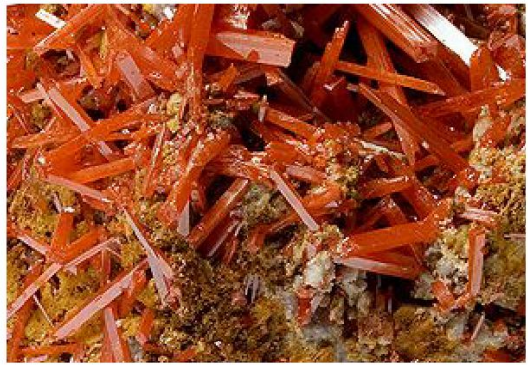

**Figure 2**

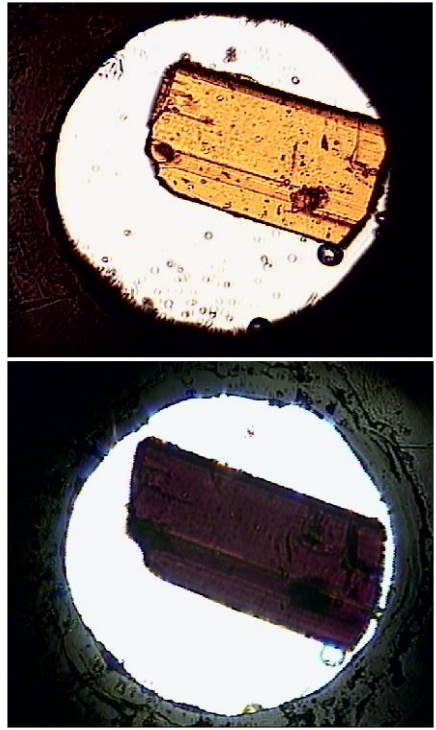

**Figure 3**

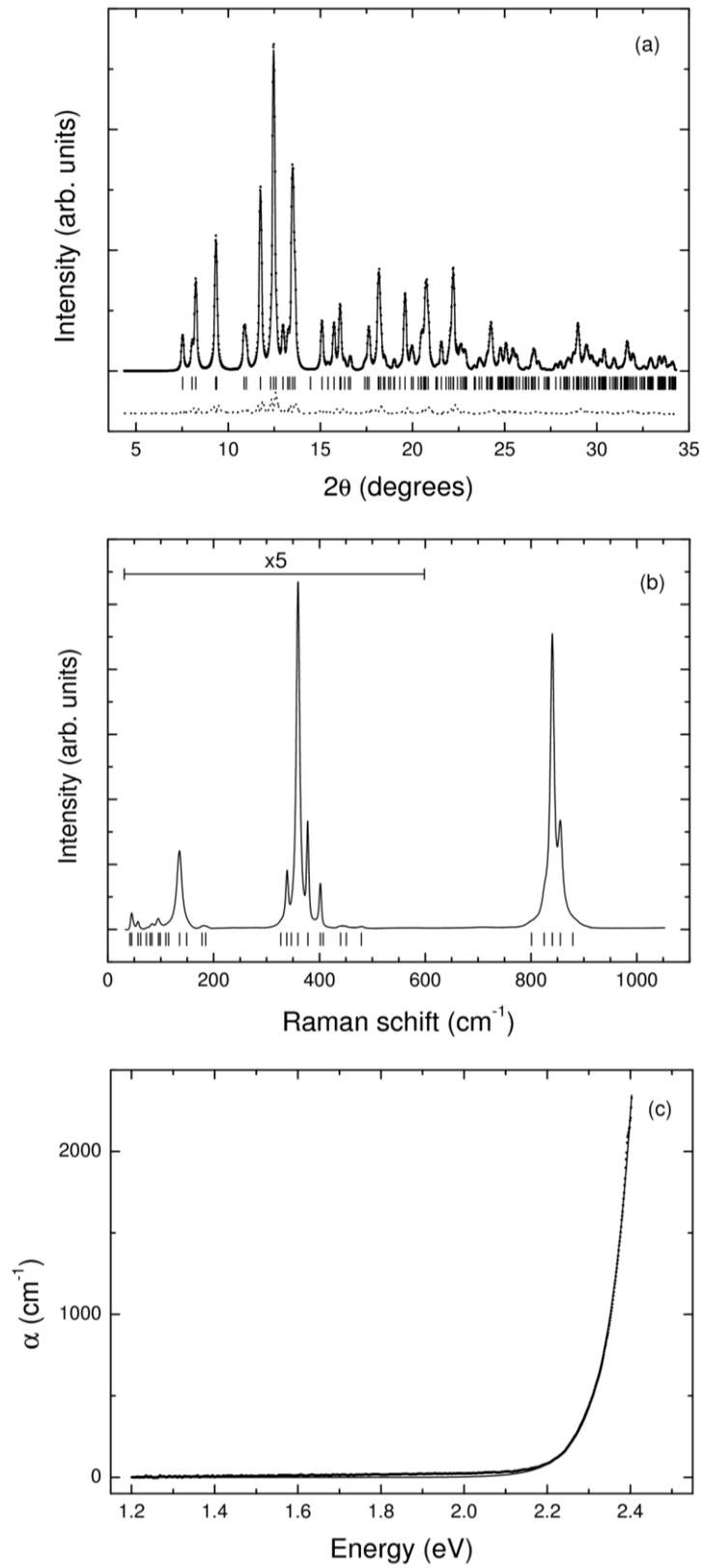

**Figure 4**

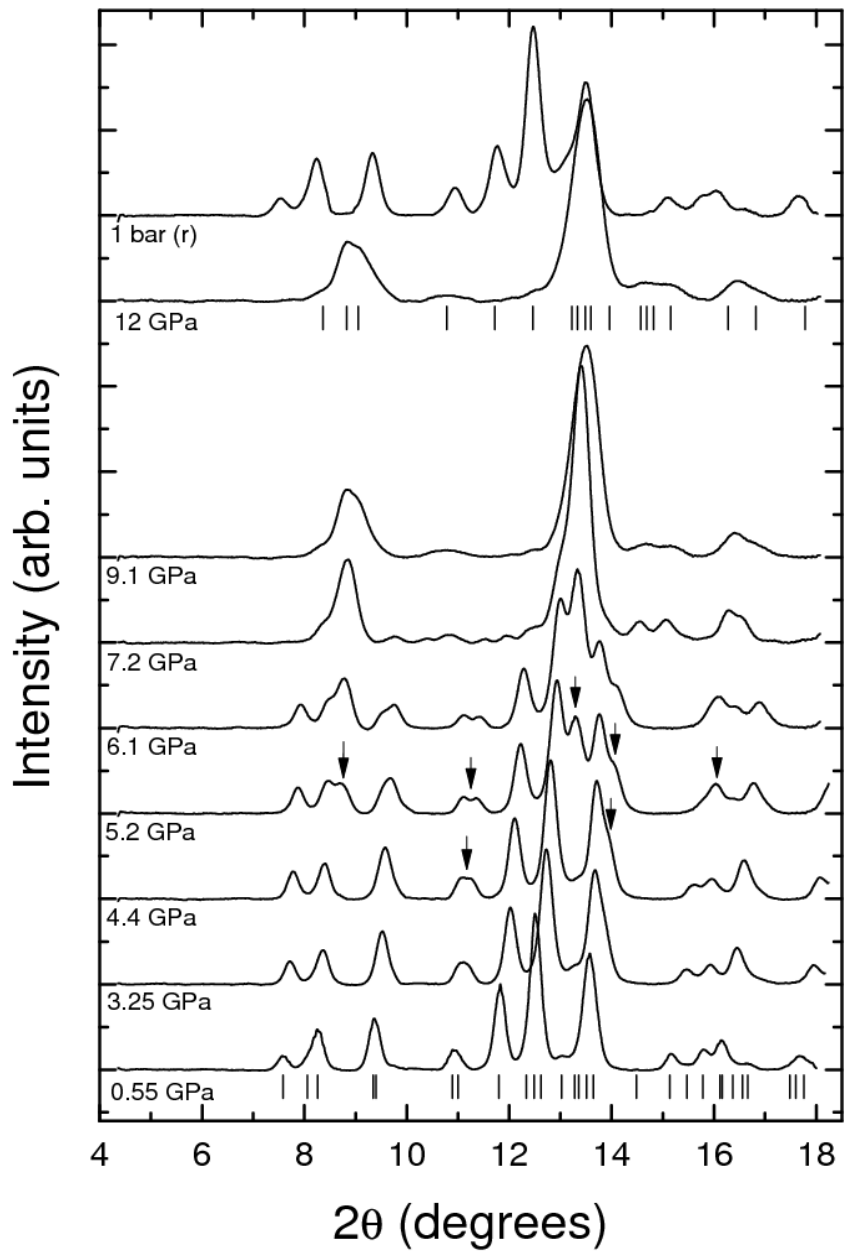

**Figure 5**

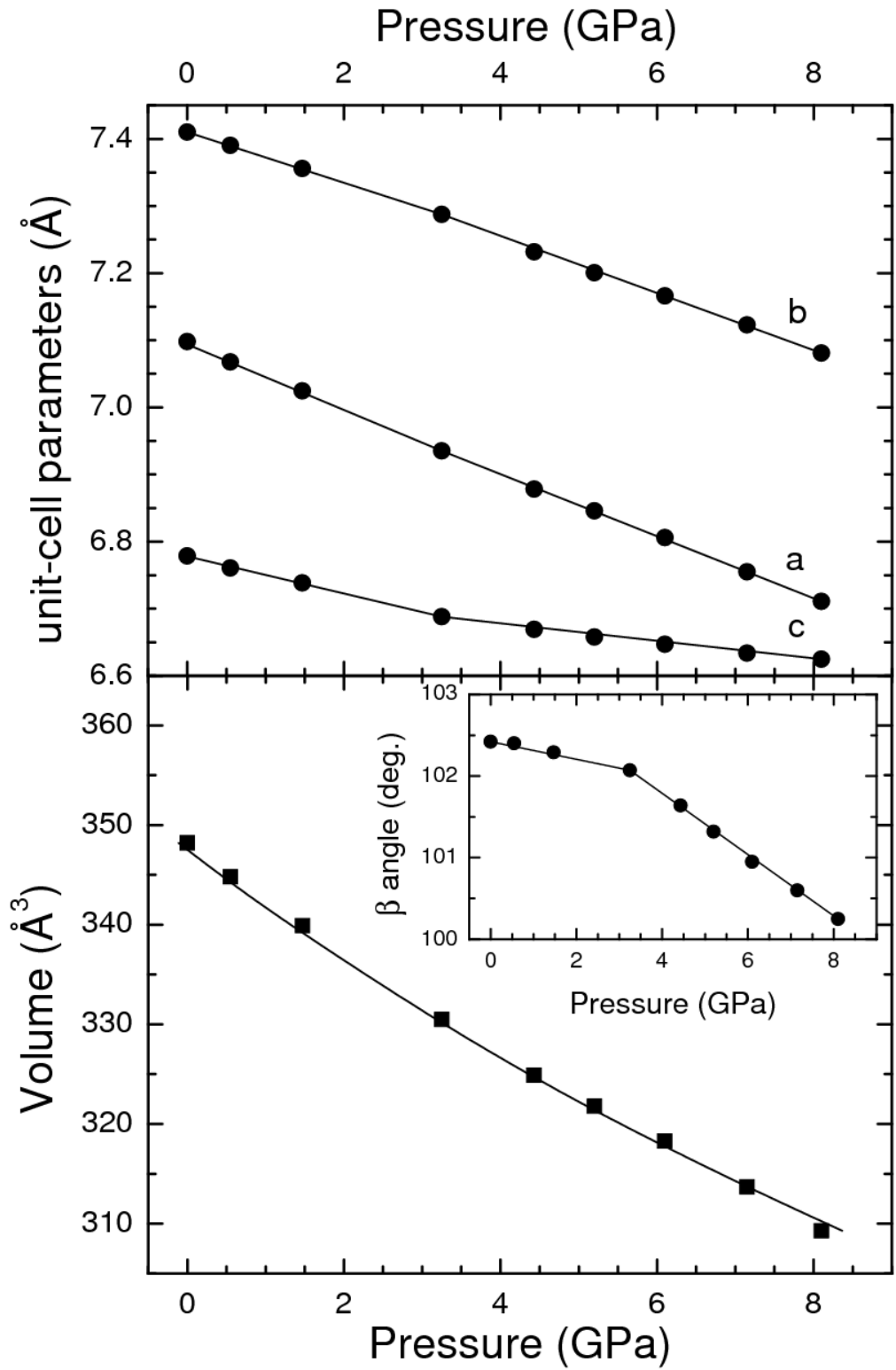

**Figure 6**

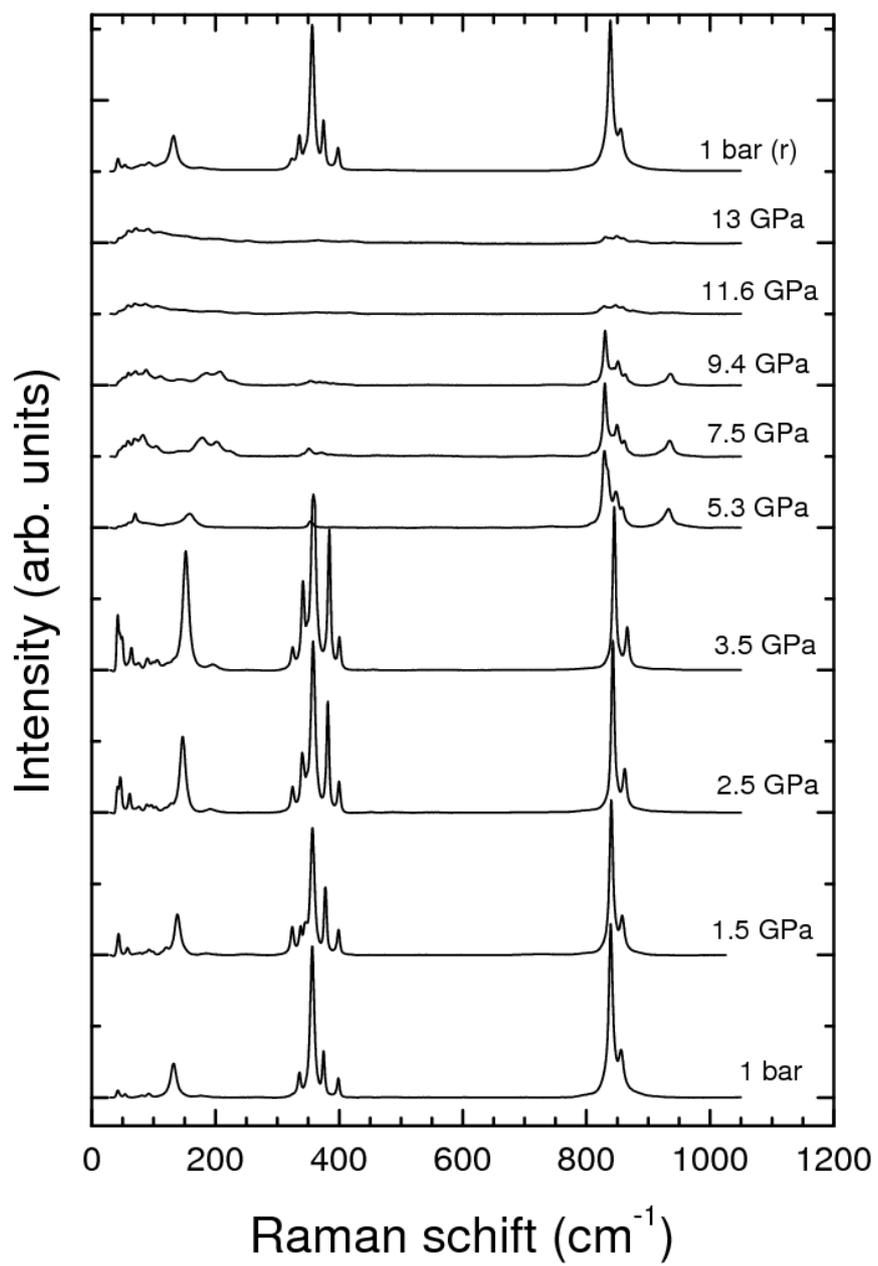

**Figure 7**

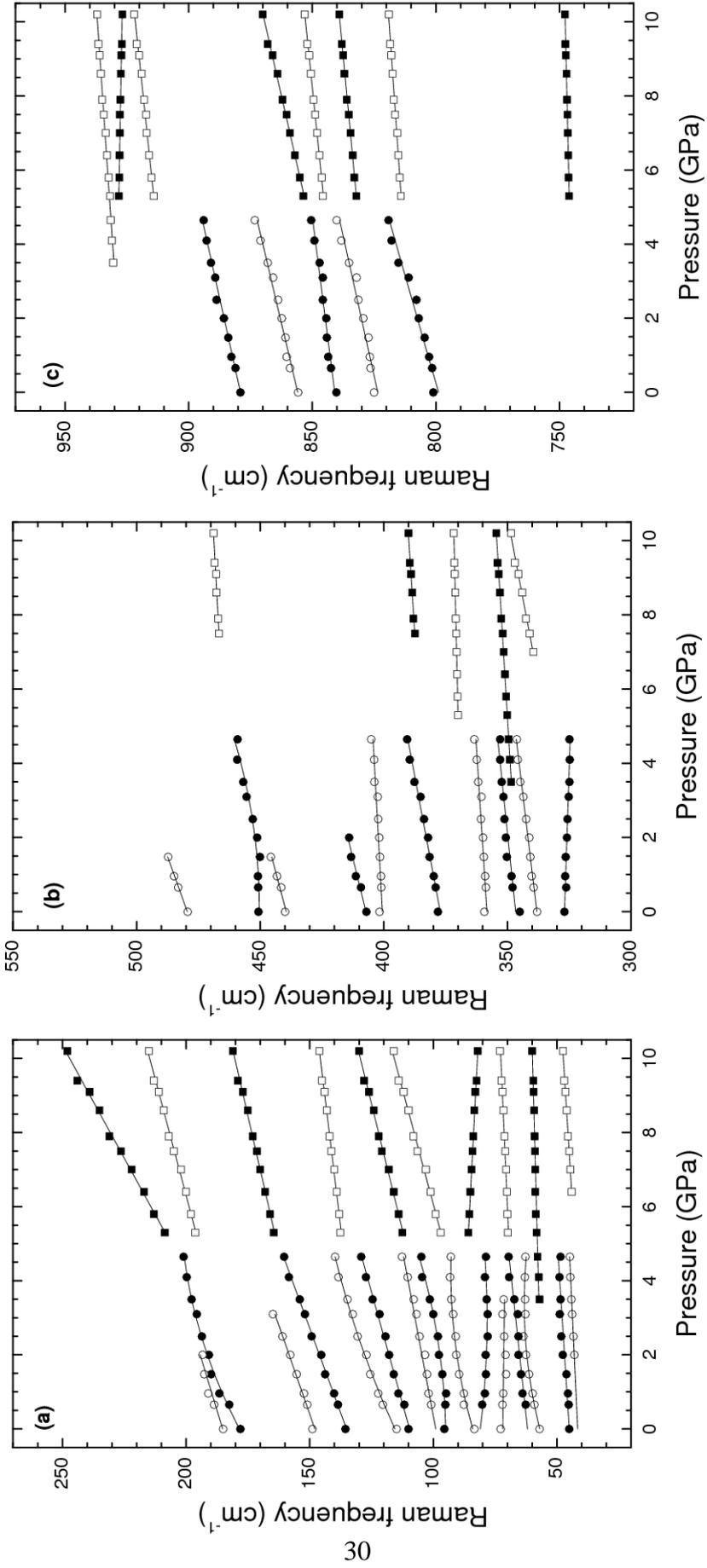

**Figure 8**

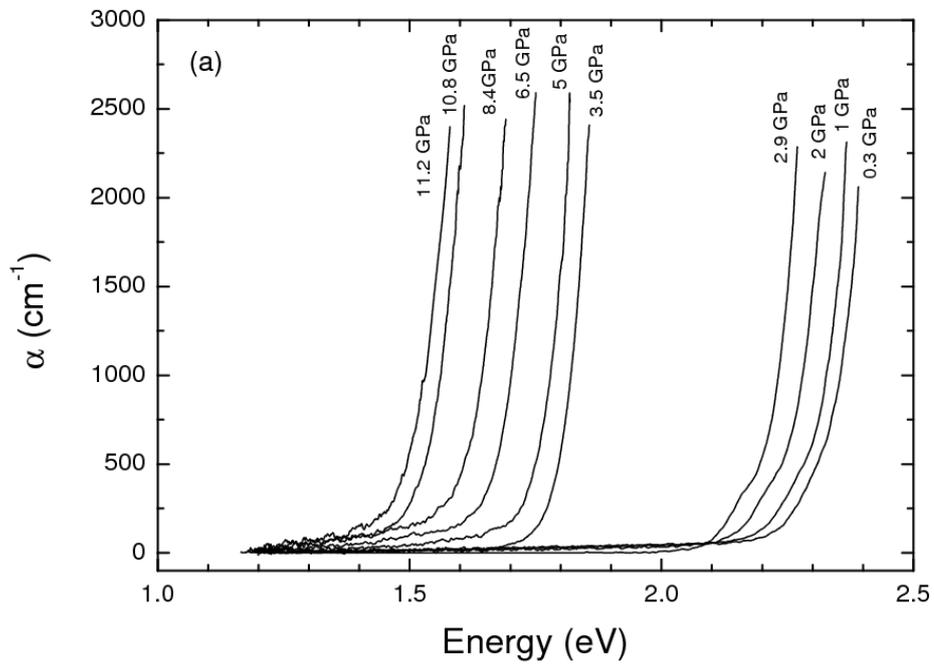

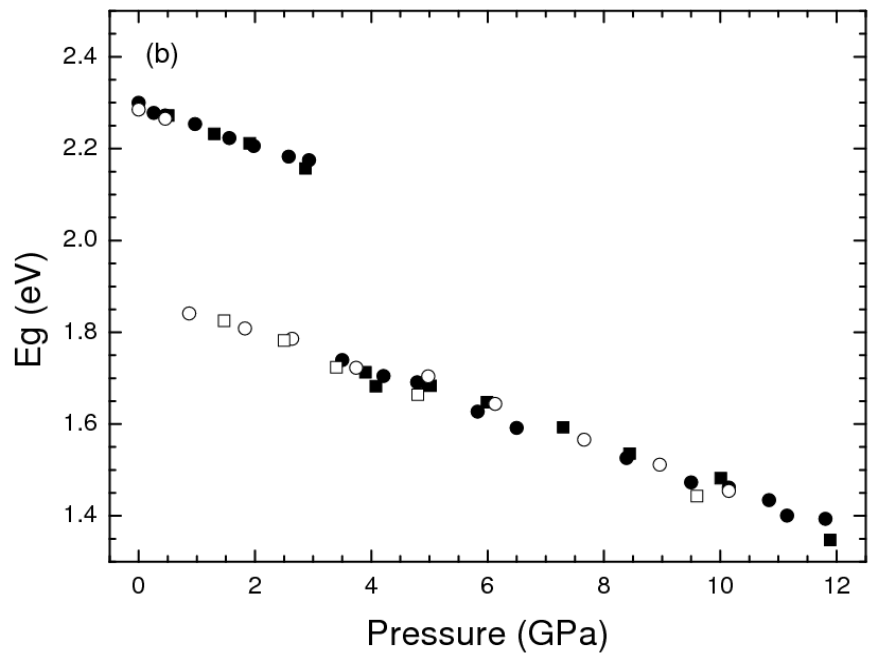

**Figure 9**

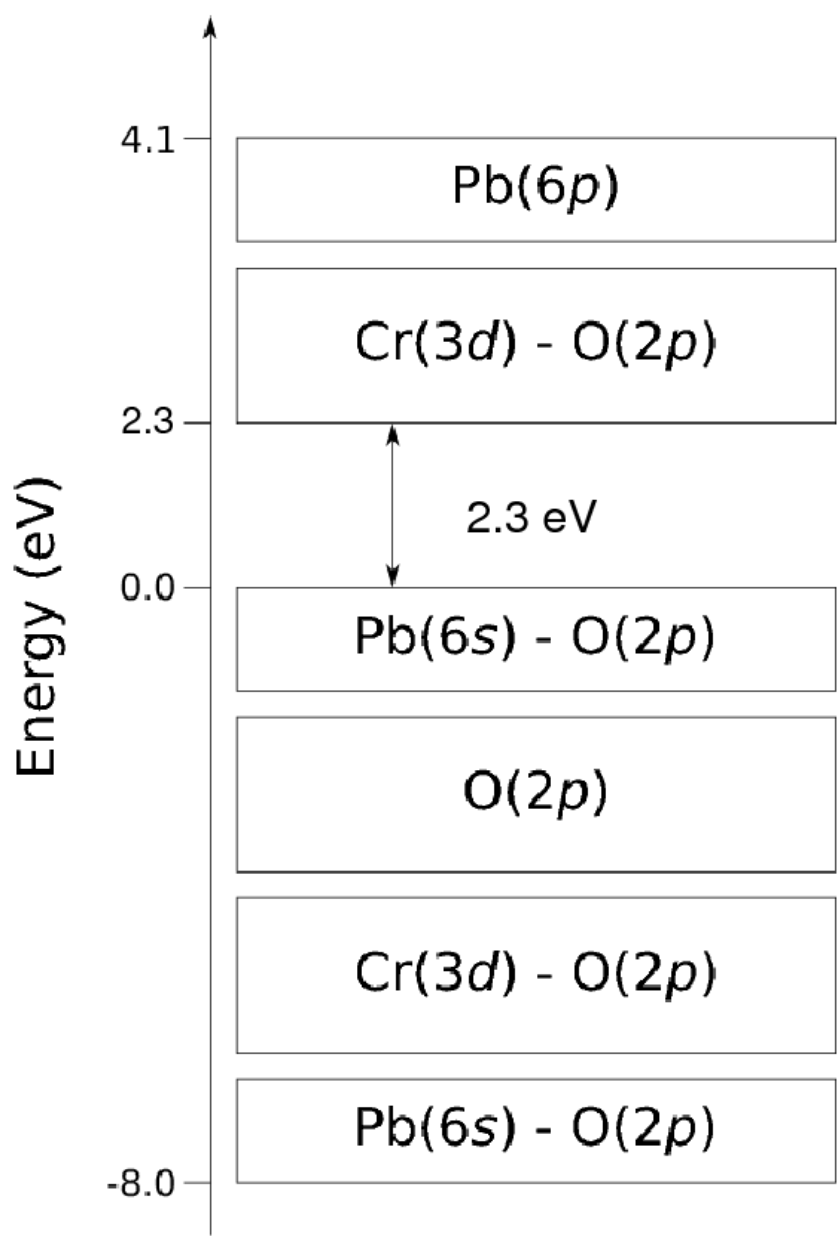